\documentclass[preprintnumbers]{revtex4}
\UseRawInputEncoding
\usepackage{amssymb}
\usepackage{amsmath}
\usepackage{graphicx}
\usepackage{dcolumn}
\usepackage{bm}
\usepackage{subfigure}
\usepackage{color}
\usepackage[latin1]{inputenc}

\begin{document}

\title{Topology of the charged AdS black hole in restricted phase space}
\author{Han Wang$^{1}$, Yun-Zhi Du$^{2,3}$\footnote{the corresponding author}}
\address{$^1$College of Mathematics and Statistics, Shanxi Datong University, Datong 037009, China\\
$^2$Department of Physics, Shanxi Datong University, Datong 037009, China\\
$^3$Institute of Theoretical Physics, Shanxi Datong University, Datong, 037009, China}

\thanks{\emph{e-mail:03090137@sxdtdx.edu.cn,duyzh22@sxdtdx.edu.cn}}

\begin{abstract}
The local topological properties of black hole systems can be expressed by the winding numbers as the defects. As so far, AdS black hole thermodynamics is often depicted by the dual parameters of $(T,S),~ (P,V), (\Phi, Q)$ in the extended phase space, while there is several study on the black hole thermodynamics in the restricted phase space. In this paper, we analyze the topological properties of the charged AdS black holes in the restricted phase space under the higher dimensions and higher order curvature gravities frame. The results show that the topological number of the charged black hole in the same canonical ensembles is a constant and is independent of the concrete dual thermodynamical parameters. However, the topological number in the grand canonical ensemble is different from that in the canonical ensemble for the same black hole system. Furthermore, these results are independent of the dimension $d$, the highest order $k$ of the Lanczos-Lovelock densities.
\end{abstract}

\maketitle

\section{Introduction}
As we know that after the last century black holes have been proposed to be not only the strong gravitational systems, but also the thermodynamic systems that satisfying the four laws of thermodynamics \cite{Bardeen1973,Bekenstein1973}. In 1983, Hawking and Page \cite{Hawking1983} proved a phase transition between a pure AdS spacetime and a stable large black hole state, i.e., Hawking-Page phase transition. Subsequently, it was explained to a confinement/deconfinement phase transition in the gauge theory \cite{Witten1998}. That leads to more attention being paid on black hole. Because of the lack of pressure in the traditional black hole thermodynamics, the authors of \cite{Kastor2009} expanded black hole thermodynamics into the expanded phase space by regarding the negative cosmological constant as pressure. Especially the thermodynamical properties of AdS black holes in the expanded phase space had been widely investigated \cite{Hendi2017a,Hennigar2017a,Frassin,Kubiznak2012,Cai2013,Ma2017,Banerjee2017,Mann1207,Wei2015,Bhattacharya2017,Zeng2017,Zhang1502,Du2021,Zhang2020}. Recently, the holographic thermodynamics \cite{Visser2022,Ahmed2023} and the restricted phase space thermodynamics \cite{Zhao2022,Gao2022,Gao2022a,Du2023} of AdS black holes had been proposed to give a holographic interpretation of black hole thermodynamics. On the other hand, based on the $\phi$-map topological flow theory \cite{Duan1979} the authors of \cite{Wei2022} presented that black hole solutions are just the defects that described by winding numbers. The corresponding winding number of the local stable black hole is one, while it is negative one for the local unstable black hole. The topological number is just the sum of all winding numbers to reveal the global topological nature of black hole. The black hole solutions may be classified by topological numbers. Based on these results, we will investigate the topological property of the charged AdS black holes under different gravity frames in the restricted phase space.

Recently, the authors of \cite{Kong2022} had checked out the restricted phase space thermodynamics in the higher dimensions and higher order curvature gravities \cite{Cai2002,Lanczos1932,Lovelock1971}. The subclass of Lanczos-Lovelock models with some particular choices of the coupling coefficients that known as the class of black hole scan models was adopted to be a simple example for the application of the extended phase space thermodynamics to the higher dimensions and higher order curvature gravities models with the character parameters of ($d,k$). The parameter $d$ stands for the spacetime dimension and $k$ is the integer with the condition $1\leq k\leq [(d-1)/2]$ that represents the highest order of the Lanczos-Lovelock densities appearing in the action. In the restricted phase space, the thermodynamics of three typical models with $(d,k)=(5,1),~(5,2),~(6,2)$ which are representative of the Einstein-Hilert (EH), Chern-Simons (CS) and Born-Infield (BI) like gravity models, were investigated. From the thermodynamical results, it is proofed that the EH and BI like models seemly belong to the same universality class while the CS like models do not. On the other hand, RN-AdS black hole is the typical solution in the EH gravity model. Thence, in this work we will probe the topology of the RN-AdS black holes and check out whether the EH and BI like models belong to the same class from the perspective of the topology in the different ensembles.

Inspired by these, firstly we investigate the topology of the RN-AdS black hole under the restricted phase space frame in two different ensembles in Sec. \ref{section2}. Then from the perspective of the topology we check out whether two typical models with $(d,k)=(5,1),~(6,2)$ belong to the same class in two different ensembles. A brief summary is given in Sec. \ref{section4}.

\section{Topology of RN-AdS black hole in the restricted phase space}
\label{section2}
In this part we will discuss the topology of the RN-AdS black hole in different ensembles under the restricted phase space.

\subsection{In the canonical ensemble}
\label{section2.1}

For the four-dimensional RN-AdS black hole, its metric has the following form
\begin{eqnarray}
ds^2=-f(r)c^2dt^2+f^{-1}(r)dr^2+r^2\left(d\theta^2+\sin^2\theta d\phi^2\right),\\
h(r)=1-\frac{2GM}{r}+\frac{G\bar q^2}{r^2}+\frac{r^2}{l^2},~~A^\mu=(\bar\phi(r)/c,0,0,0),~~
\bar\Phi(r)=\frac{\bar q}{r},
\end{eqnarray}
where $l$ is related to the cosmological constant via $l^2=-3/\Lambda$. As $M>\bar q/\sqrt{G}$, the function $h(r)$ has two distinct real zeros at $r=r_{\pm}$, $r_+$ corresponds to the black hole event horizon radius. The black hole mass parameter can be described as
\begin{eqnarray}
M=\frac{r_+}{2G}\left(1+\frac{G\bar q^2}{r_+^2}+\frac{r_+^2}{l^2}\right).
\end{eqnarray}
In the restricted phase space for the RN-AdS black hole, the macro states are characterized by the following three pairs of dual thermodynamical parameters $(S,T),~(Q,\Phi),$ and $(\mu,N)$. The corresponding first law of thermodynamics and smarr formula read
\begin{eqnarray}
dM=TdS+\Phi dQ+\mu dN,~~M=TS+\Phi Q+\mu N
\end{eqnarray}
with
\begin{eqnarray}
M(S,Q,N)&=&\frac{S^2+\pi SN+\pi^2Q^2}{2\pi^{3/2}l(SN)^{1/2}},~~T(S,Q,N)=\frac{3S^2+\pi SN-\pi^2Q^2}{4\pi^{3/2}lS(SN)^{1/2}},\\
\Phi&=&\left(\frac{\pi}{SN}\right)^{1/2},~~\mu=-\frac{S^2-\pi SN+\pi^2Q^2}{4\pi^{3/2}lN(SN)^{1/2}},\\
G&=&\frac{l^2}{N},~~S=\frac{\pi r_+^2}{G},~~Q=\frac{ql}{\sqrt{G}}.\label{GSQ}
\end{eqnarray}
If the parameters $S,~Q,~N$ are scaled as $S\rightarrow\lambda S,~Q\rightarrow\lambda Q,~N\rightarrow\lambda N$, from eq. (\ref{GSQ}) we can see that the mass parameter will also be scaled as $M\rightarrow\lambda M$, and $T,~\Phi,~\mu$ will not be scaled. Hence the first order homogeneity of the mass parameter and the zeroth order homogeneity of $T,~\Phi,~\mu$ are clear as crystal. Note that the zeroth order homogeneous functions are intensive the critical point
\begin{eqnarray}
S_c=\frac{\pi N}{6},~~T_c=\frac{\sqrt{6}}{3\pi l},~~Q_c=\frac{N}{6},~~F_c=\frac{\sqrt{6}N}{18l}.
\end{eqnarray}
Introducing the relative parameters $s=S/S_c,~t=T/T_c,~q=Q/Q_c$ and a Legendre transform of eq. (\ref{GSQ}), the relative parameter of free energy is given as \cite{Gao2022a}
\begin{eqnarray}
f\equiv\frac{F}{F_c}=\frac{q^2+s^2+6s-4ts^{3/2}}{4s^{1/2}},
~~F(T,Q,N)=M(T,Q,N)-TS. \label{fRS}
\end{eqnarray}

In order to uncover the thermodynamical topology, the vector field mapping $\phi:~X=\{(s,\theta)|0<s<\infty,$ $0<\theta<\pi\}\rightarrow \mathbb{R}^2$ is defined as \cite{Wei2020,Wei2022,Wei2022a,Wu2023,Wu2023a}
\begin{eqnarray}
\phi(s,\theta)=\left(\frac{\partial f}{\partial s},-\cot\theta\csc\theta\right).\label{phi}
\end{eqnarray}
When the parameter $s$ is used to characterize AdS black hole, it becomes the first parameter of the domain of the mapping $\phi$. Another parameter $\theta$ serves as an auxiliary function and it is utilized to construct the second component of the mapping. The component $\phi^\theta$ is divergent at $\theta=0,~\pi$, thus the direction of the vector points outward there. It is obvious that the zero point of $\phi$ corresponds to the black hole with the temperature of $T=\tau^{-1}$ as $\theta=\pi/2$. Therefore, the zero point of the mapping can be used to characterize the black hole solution with a given parameter $\tau$. Based on the Duan's $\phi-$mapping topological current theory \cite{Duan1979,Duan1984}, the zero points of the mapping $\phi$ is linked to the topological number. The topological number can be obtained by the weighted sum of the zero points of the mapping $\phi$. The weight of each zero point is determined by its nature. A saddle point has a weight of negative one, and conversely, the weight of an extremum point is one. A topological current can be described as the following form
\begin{eqnarray}
j^\mu=\frac{1}{2}e^{\mu\nu\rho}\epsilon_{ab}\partial_\nu n^a \partial_\rho n^b,~~~~\mu,\nu,\rho=0,1,2,~~a,b=1,2,
\end{eqnarray}
where $n$ is the unit vector $(n^s,n^\theta)$ with $n^s=\phi^s/\parallel\phi\parallel$ and $n^\theta=\phi^\theta/\parallel\phi\parallel$. And it satisfies the conserve law: $\partial_\mu j^\mu=0$. As shown in Refs. \cite{Duan1998,Fu2000}, the topological current is a $\delta$-function of the field configuration
\begin{eqnarray}
j^\mu=\delta^2(\phi)J^\mu\left(\frac{\phi}{x}\right),
\end{eqnarray}
where the three-dimensional Jacobian reads $\epsilon^{ab}J^\mu\left(\frac{\phi}{x}\right)=\epsilon^{\mu\nu\rho}\partial_\nu \phi^a \partial_\rho \phi^b$. As $\phi^a(x_i)=0$, the topological current equals to zero. The topological number in a parameter region $\Sigma$ can be calculated by the following expression
\begin{eqnarray}
W=\int_\Sigma j^0d^2x=\sum\limits^{\bar N}_{i=1}\beta_i\eta_i=\sum\limits^{\bar N}_{i=1}w_i,
\end{eqnarray}
where $j^0=\sum\limits^{\bar N}_{i=1}\beta_i\eta_i\delta^2\left(\overrightarrow{x}-\overrightarrow{s}_i\right)$
is the density of the topological current, $\beta_i$ is the Hopf index which is always positive. $s_i$ is the $i$-th zero point of the mapping $\phi$, $\eta_i=sign J^0(\phi/x)_{s_i}=\pm 1$ is the Brouwer degree. For the $i$-th zero point of the vector, the winding number $w_i$ is determined by the stability of black hole state.

\begin{figure}[htp]
\centering
\includegraphics[width=0.45\textwidth]{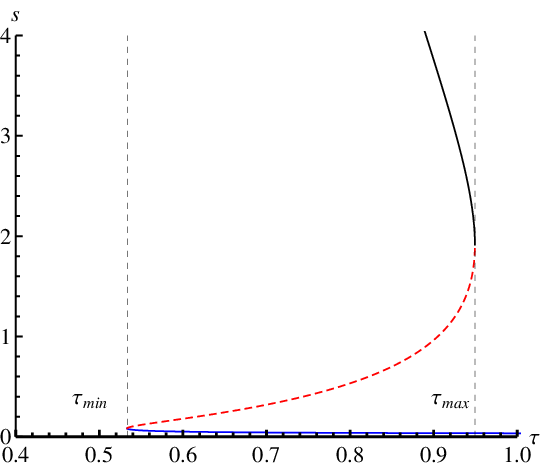}
\caption{Zero points of $\phi^s$ in the diagram of $\tau-s$ with $q=0.4$ for the RN-AdS black hole in the canonical ensemble.}\label{taus}
\end{figure}

In the canonical ensemble, substituting eq. (\ref{fRS}) into eq. (\ref{phi}), we have
\begin{eqnarray}
\phi^s=\frac{\partial f}{\partial s}=-\frac{q^2}{8s^{3/2}}+\frac{3s^{1/2}}{8}+\frac{3}{4s^{1/2}}-\frac{1}{\tau},
~~\phi^\theta=-\cot\theta\csc\theta.
\end{eqnarray}
The zero points are determined by the expression of $\phi^s=0$. So we have
\begin{eqnarray}
\frac{1}{\tau}=-\frac{q^2}{8s^{3/2}}+\frac{3s^{1/2}}{8}+\frac{3}{4s^{1/2}}.
\end{eqnarray}
It is obvious that there are two extremes: $\tau_{min,max}=\frac{2\left(1\mp\sqrt{1-q^2}\right)^{3/2}}{3(1\mp\sqrt{1-q^2})-q^2}$.
As $q=1$, $\tau_{min}=\tau_{max}=\tau_c=1$. Note that the generation point satisfies the constraint conditions: $\frac{\partial\tau}{\partial s}=0,~\frac{\partial^2\tau}{\partial s^2}>0$, and the annihilation point obeys the conditions: $\frac{\partial\tau}{\partial s}=0,~\frac{\partial^2\tau}{\partial s^2}<0$. The zero points of $\phi^s$ in the diagram of $\tau-s$ is displayed in Fig. \ref{taus}. $\tau_{min}$ is related with the generation point, and $\tau_{max}$ is of the annihilation point. As $\tau_{min}<\tau<\tau_{max}$, there are three intersection points for the RN-AdS black hole in the canonical ensemble. The black, red dashed, blue lines are for the low-potential black hole (LPSB), intermediate-potential black hole (IPBH), and high-potential black hole (HPBH), respectively. They are just the stable, unstable, and stable black hole states. The corresponding winding numbers are 1, -1, 1, so the topological number is $W=1-1+1=1$ as $\tau_{min}<\tau<\tau_{max}$. The intersection points exactly are satisfying the equation $\tau=1/T$. For $\tau=\tau_{min,max}$, the three intersection points for the RN black hole coincide. And they will disappear as $\tau<\tau_{min}$ or $\tau>\tau_{max}$, the three branches black holes reduce to the one stable black hole and the topological number is still one. These results are consistent with that in Ref. \cite{Fan2023}.

\subsection{In the grand canonical ensemble}
\label{section2.2}

\begin{figure}[htp]
\centering
\includegraphics[width=0.45\textwidth]{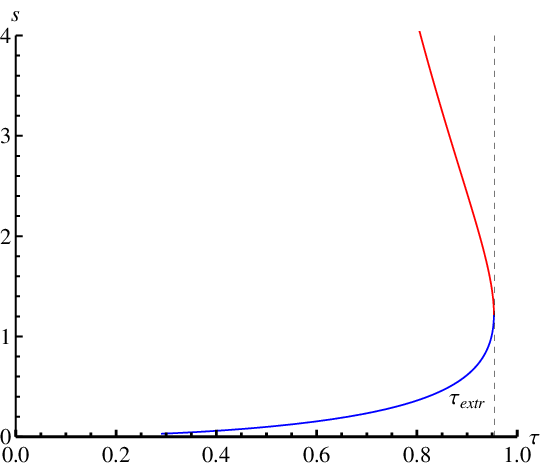}
\caption{Zero points of $\phi^s$ in the diagram of $\tau-s$ with $l^2\Phi^2=0.1$ for the RN-AdS black hole in the grand canonical ensemble.}\label{taus1}
\end{figure}

Now we proceed to investigate the topology of the RN-AdS black hole in the grand canonical ensemble. For the given values of $\Phi$ and $N$, from eq. (\ref{GSQ}) the minimum temperature reads \cite{Gao2022a}
\begin{eqnarray}
T_{min}=\frac{\sqrt{3}}{2}\frac{\left(1-l^2\Phi^2\right)^{1/2}}{\pi l},
\end{eqnarray}
the corresponding minimum entropy, mass parameter, and free energy are
\begin{eqnarray}
S_{min}=\frac{\pi N}{3}\left(1-l^2\Phi^2\right),~~
M_{min}=\frac{N\left(2+l^2\Phi^2\right)}{3\sqrt{3}l}\left(1-l^2\Phi^2\right)^{1/2},~~
F_{min}=\frac{N\left(1+5l^2\Phi^2\right)}{6\sqrt{3}l}\left(1-l^2\Phi^2\right)^{1/2}.
\end{eqnarray}
By introducing the relative parameters $t=T/T_{min},~s=S/S_{min},~m=M/M_{min},~f=F/F_{min}$, we can obtain the express
\begin{eqnarray}
f=\frac{s^{1/2}\left[3+s+(3-s)l^2\Phi^2\right]}{1+5l^2\Phi^2}-\frac{3ts\left(1-l^2\Phi^2\right)^{1/2}}{1+5l^2\Phi^2},
\end{eqnarray}
so the component of the mapping $\phi^s$ reads
\begin{eqnarray}
\phi^s=\frac{3\left[1+s(1-l^2\Phi^2)+l^2\Phi^2\right]}
{2s^{1/2}(1+5l^2\Phi^2)}-\frac{3t\left(1-l^2\Phi^2\right)^{1/2}}{1+5l^2\Phi^2}.
\end{eqnarray}
Hence, we have
\begin{eqnarray}
\tau=\frac{2s^{1/2}\left(1-l^2\Phi^2\right)^{1/2}}{\left[1+s(1-l^2\Phi^2)+l^2\Phi^2\right]}.
\end{eqnarray}
There exist one extreme: $\tau_{extr}=\sqrt{\frac{\left(1-l^4\Phi^4\right)}{\left(1+l^4\Phi^4\right)}}$.
The zero points of $\phi^s$ in the diagram of $\tau-s$ is displayed in Fig. \ref{taus1}. $\tau_{extr}$ is related with the annihilation point. As $\tau<\tau_{extr}$, there are two intersection points for the RN-AdS black hole in the grand canonical ensemble. The red and blue lines are for the low-potential black hole (LPSB) and high-potential black hole (HPBH), respectively. They are just the stable and unstable black hole states. The corresponding winding numbers are 1, -1, so the topological number is $W=1-1=0$ as $\tau<\tau_{extr}$. For $\tau=\tau_{extr}$, the two intersection points for the RN black hole coincide. Furthermore, there is no black hole as $\tau>\tau_{extr}$.

\section{Topology of the AdS black hole in the higher dimensional and higher curvature gravity}
\label{section3}

Recently, the restricted phase space thermodynamics is shown to be applicable to a large class of the higher dimensional and higher curvature gravity models with the coupling to the Maxwell field \cite{Kong2022,Gogoi2023}, which are known as black hole scan models \cite{Crisostomo2000} and are labeled by the spacetime dimension $d$ and the highest order $k$ of the Lanczos-Lovelock densities appearing in the action. Three typical example cases with $(d, k) = (5, 1)$, $(5, 2)$, and $(6, 2)$ are chosen as example cases and studied in some detail. These cases are representatives of Einstein-Hilbert, Chern-Simons and Born-Infield like gravity models. However, the Chern-Simons like $(5, 2)$- model behaves quite differently. This seems to indicate that the Einstein-Hilbert and Born-Infield like models belong to the same universality class while the Chern-Simons like models do not. In this part we will check out whether the EH and BI like models do belong the same universality class from topology.

To be more concrete, the metric function of the black holes in the higher dimensions and higher curvature gravities with the coupling to Maxwell's field was given in Refs.
\cite{Kong2022,Gogoi2023,Crisostomo2000}
\begin{eqnarray}
ds^2_{(d,k)}&=&-f_{(d,k)}(r)dt^2+f_{(d,k)}^{-1}(r)dr^2+r^2d\Omega^2_{d-2},\nonumber\\
f_{(d,k)}(r)&=&1+\frac{r^2}{l^2}-\left(\frac{2G_{(d,k)}M+\delta_{d-2,k,1}}{r^{d-2k-1}}
-\frac{G_{(d,k)}Q^2}{(d-3)r^{2(d-k-2)}}\right)^{1/k},~~\Phi=\frac{Q}{(d-3)r^{d-3}},\label{fdk}
\end{eqnarray}
where $G_{(d,k)}$ is the Newton constant, and $\delta_{d-2,k,1}$ is the Kronecker delta function. $Q$ and $M$ are the charge and mass parameters of black hole. Note that for $k=1$, the three dimensional black hole \cite{Banados1992,Banados1993} and Schwarzschild-AdS solutions of the higher dimensional Einstein-Hilbert action with the negative cosmological constant both can be recovered. And when setting $k=\left[\frac{d-1}{2}\right]$, we also can obtain the black hole solutions corresponding to Born-Infeld (BI) and Chern-Simons (CS) theories \cite{Banados1994}.

The black hole horizon $r_h$ is located at one of the zero points of $f_{(d,k)}(r)$. By solving the equation of $f_{(d,k)}(r_h)=0$, the black hole mass parameter can be obtained as the following form
\begin{eqnarray}
M(r_h,G_{(d,k)},Q)=-\frac{\delta_{d-2,k,1}}{2G_{(d,k)}}
+\frac{r_h^{d-2k-1}\left(1+r_h^2/l^2\right)^k}{2G_{(d,k)}}
+\frac{Q^2}{2(d-3)r_h^{d-3}}.
\end{eqnarray}
The temperature of the black hole can be evaluated by the Euclidean period method, which gives
\begin{eqnarray}
T=\frac{1}{4\pi k r_h^{2d-1}}\left(1+r_h^2/l^2\right)^{1-k}
\left[(d-2k-1)r_h^{2d}\left(1+r_h^2/l^2\right)^{k}-G_{(d,k)}Q^2r_h^{2k+4}\right]
+\frac{r_h}{2\pi l^2}.\label{Tdk}
\end{eqnarray}
The entropy of the black hole was given in Ref. \cite{Gogoi2023} as the following form
\begin{eqnarray}
S=\frac{2\pi k}{G_{(d,k)}}\int_0^{r_h}r^{d-2k+1}
\left(1+r^2/l^2\right)^{k-1}dr.\label{Sdk}
\end{eqnarray}
For establishing the restricted phase space thermodynamics (RPST) of the $(d,k)-$models, we introduce two thermodynamical parameters: the effective number of the microscopic degrees of freedom $N$ and the chemical potential $\mu$, which are defined as
\begin{eqnarray}
N=\frac{L^{d-2k}}{G_{(d,k)}},~~\mu=(M-TS-\Phi Q)/N.\label{Nmu51}
\end{eqnarray}
Here, $L$ is a constant that is introduced to make $N$ dimensionless. The corresponding re-scaled charge and potential are
\begin{eqnarray}
\widetilde{Q}=\frac{QL^{(d-2k)/2}}{\sqrt {G_{(d,k)}}},~~
\widetilde{\Phi}=\frac{Q\sqrt {G_{(d,k)}}}{(d-3)L^{(d-2k)/2}r_h^{d-3}}{\color{red}.}\label{QPhi51}
\end{eqnarray}
The thermodynamical parameters are satisfied the following expressions
\begin{eqnarray}
M=TS+\widetilde{\Phi}\widetilde{Q}+\mu N,~~dM=TdS+\widetilde{\Phi}d\widetilde{Q}+\mu dN.
\end{eqnarray}
Note that the above equations indicate that the RPST formalism holds on for the $(d,k)-$models with $d>3$.

\subsection{Topology of RPST for $(5,1)$-model: EH gravity coupled to Maxwell's field in the canonical ensemble}
\label{section3.1}

Setting $(d,k)=(5,1)$, the eq. (\ref{fdk}) becomes
\begin{eqnarray}
f_{(5,1)}=1+\frac{r^2}{l^2}+\frac{G_{(5,1)}Q^2}{2 r^4}
-\frac{2G_{(5,1)}M}{r^2}.
\end{eqnarray}
This is the black hole solution in the Einstein-Hilbert gravity theory, and the entropy is $S=A/4G$ with the Newton constant $G=3\pi G_{(5,1)}/4$. Through the transformations
\begin{eqnarray}
r_h=[3SL^3/(2\pi N)]^{1/3},~G_{(5,1)}=L^3/N,~Q=\widetilde{Q}/\sqrt{N},
~\widetilde{S}=(3SL^3/N)^{1/3},\label{trans51}
\end{eqnarray}
the temperature and mass parameter of the black hole can be rewritten as
\begin{eqnarray}
T&=&\frac{\widetilde{S}}{2(2\pi)^{2/3}}
\left(2+4\widetilde{S}^2(2\pi)^{-2/3}/l^2
-\frac{(2\pi)^{4/3}L^3\widetilde{Q}^2}{N^2\widetilde{S}^4}\right),\label{T51}\\
M&=&\frac{\widetilde{S}^2N}{2(2\pi)^{2/3}L^3}
\left(1+\frac{(2\pi)^{-2/3}\widetilde{S}^2}{l^2}
+\frac{(2\pi)^{4/3}L^3\widetilde{Q}^2}{2N^2\widetilde{S}^4}\right).\label{M51}
\end{eqnarray}
It is transparent that when $\widetilde{S}\rightarrow\lambda\widetilde{S},
~\widetilde{Q}\rightarrow\lambda\widetilde{Q},~N\rightarrow\lambda N$, the mass parameters behaves as $M\rightarrow\lambda M$, while the temperature is kept unchanged under these re-scalings. From these thermodynamical parameters, we can obtain the critical point
\begin{eqnarray}
\widetilde{S}_c=\frac{(2\pi)^{1/3}l}{\sqrt{3}},~~
\widetilde{Q}_c=\sqrt{\frac{2}{135}}\frac{l^2N}{L^{3/2}},~~
T_c=\frac{4\sqrt{3}}{5\pi l},~~
M_c=\frac{7l^2N}{30L^3},~~ F_c=\frac{l^2N}{18L^3}.
\end{eqnarray}
\begin{figure}[htp]
\centering
\includegraphics[width=0.45\textwidth]{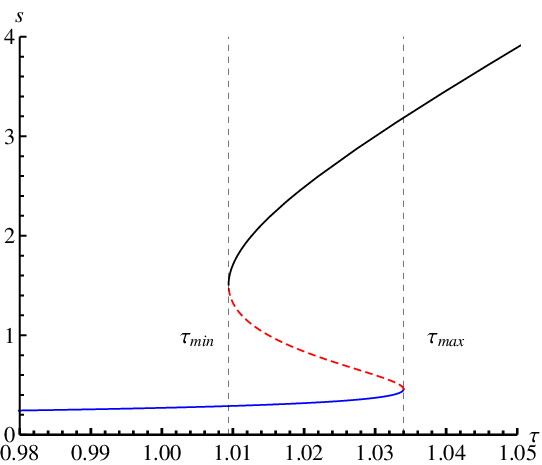}
\caption{Zero points of $\phi^s$ in the diagram of $\tau-s$ with $q=0.8$ for the higher dimensions and higher order curvature gravity model with $(d,k)=(5,1)$ in the canonical ensemble.}\label{taus51}
\end{figure}
With the relative parameters $t=T/T_c,~s=S/S_c,~q=Q/Q_c,
~\widetilde{\phi}=\widetilde{\Phi}/\widetilde{\Phi}_c,~m=M/M_c,~f=F/F_c$, the relative free energy reads
\begin{eqnarray}
f=mM_c/F_c-tsT_cS_c/F_c=s^{4/3}+3s^{2/3}+\frac{q^2}{5s^{2/3}}-\frac{16s}{5\tau}.
\end{eqnarray}
From the above equation, the components of the mapping $\phi$ can be written as
\begin{eqnarray}
\phi^s=\frac{4s^{1/3}}{3}+\frac{2}{s^{1/3}}-\frac{2q^2}{15s^{5/3}}-\frac{16}{5\tau},~~
\phi^\theta=-\cot\theta\csc\theta.
\end{eqnarray}
The zero points of the mapping $\phi$ can be calculated by $\phi^s=0$. We solve it and get
\begin{eqnarray}
\tau=\frac{5s^{1/3}}{12}+\frac{5}{8s^{1/3}}-\frac{q^2}{24s^{5/3}}.
\end{eqnarray}
In the Fig. \ref{taus51}, we can see that there are three branches as $\tau_{min}<\tau<\tau_{max}$, the large black hole branch for
$\tau<\tau_{min}$, and the small black hole branch for $\tau>\tau_{max}$. That means the existence of a phase transition in the region of $\tau_{min}<\tau<\tau_{max}$. The winding numbers of the large black hole branch for $\tau<\tau_{min}$ and small black hole branch for $\tau>\tau_{max}$, which are stable, both read $w=1$. And the corresponding topological number equals to one. While for the system in the region of $\tau_{min}<\tau<\tau_{max}$, there exist the stable large and small black hole, and the unstable intermediate black hole, thus the corresponding winding number are $w=1,1,-1$. The topological number is $W=1$. These results are consistent with that for the RN-AdS black hole in the canonical ensemble.

\subsection{Topology of RPST for $(5,1)$-model: EH gravity coupled to Maxwell's field in the grand canonical ensemble}
\label{section3.2}

In the grand canonical ensemble, using the eqs. (\ref{Nmu51}), (\ref{QPhi51}), and (\ref{trans51}), we can obtain
\begin{eqnarray}
Q^2=\frac{4N\widetilde{\Phi}^2}{(2\pi)^{4/3}}
\left(\frac{3L^3S}{N}\right)^{4/3},~~
\widetilde{S}=L\left(\frac{3S}{N}\right)^{1/3},\label{Q2S}
\end{eqnarray}
then substituting them into eqs. (\ref{T51}) and (\ref{M51}), the temperature and mass parameter become
\begin{eqnarray}
T&=&\left(\frac{3\pi^2NL^3}{2S}\right)^{1/3}
\left(2+\frac{4L^2}{l^2}\left[\frac{3S}{2\pi N}\right]^{2/3}
-4\widetilde{\Phi}^2L^3\right),\\
M&=&\frac{N}{2L}\left(\frac{3S}{2\pi N}\right)^{2/3}
\left(1+\frac{L^2}{l^2}\left[\frac{3S}{2\pi N}\right]^{2/3}+2\widetilde{\Phi}^2L^3\right).
\end{eqnarray}
From the above equation, the minimum temperature is determined by the expression $\frac{\partial T}{\partial S}=0$, yielding
\begin{eqnarray}
T_{min}=\frac{\sqrt{2(1-2\widetilde{\Phi}^2L^3)}}{\pi l},
\end{eqnarray}
the corresponding minimum entropy, mass and free energy read
\begin{eqnarray}
S_{min}=\frac{\pi Nl^3\left(1-2\widetilde{\Phi}^2L^3\right)^{3/2}}{3\sqrt{2}L^3},~~
M_{min}=\frac{Nl^2(1-2L^3\widetilde{\Phi}^2)(3+2L^3\widetilde{\Phi}^2)}{8L^3},~~
F_{min}=\frac{N(1+5l^2\widetilde{\Phi}^2)
\left(1-\widetilde{\Phi}^2l^2\right)^{1/2}}{6\sqrt{3}l}.
\end{eqnarray}
\begin{figure}[htp]
\centering
\includegraphics[width=0.45\textwidth]{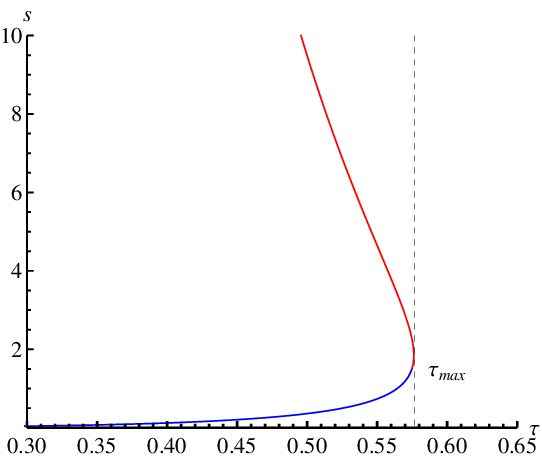}
\caption{Zero points of $\phi^s$ in the diagram of $\tau-s$ with $L^3\widetilde{\Phi}^2=0.1$ for the AdS black hole under the (5,1)-model frame in the grand canonical ensemble.}\label{taus151}
\end{figure}
With the definitions $f=F/F_{min}=(M-TS)/F_{min},~1/\tau=t=T/T_{min},~s=S/S_{min}$, we can obtain the components of the mapping $\phi$
\begin{eqnarray}
\phi^s=\frac{\partial f}{\partial s}=\frac{1}{F_{min}}
\left(\frac{\partial M}{\partial s}-\frac{T_{min}S_{min}}{\tau}\right),~~
\phi^\theta=-\cot\theta\csc\theta.
\end{eqnarray}
The zero points of the mapping $\phi$ can be calculated by $\phi^s=0$. We solve it and get
\begin{eqnarray}
\tau=\frac{\sqrt{2}s^{1/3}(1-2L^3\widetilde{\Phi}^2)}
{1+2L^3\widetilde{\Phi}^2+s^{2/3}(1-2L^3\widetilde{\Phi}^2)}.
\end{eqnarray}
As $s=(1+2L^3\widetilde{\Phi}^2)^{3/2}/(1-2L^3\widetilde{\Phi}^2)^{3/2}$, $\tau=\tau_{max}=\frac{\sqrt{2}}{
\left(\left[\frac{1+2L^3\widetilde{\Phi}^2}{1-2L^3\widetilde{\Phi}^2}\right]^{1/3}
+\left[\frac{1+2L^3\widetilde{\Phi}^2}{1-2L^3\widetilde{\Phi}^2}\right]^{2/3}\right)}$. The zero points of $\phi^s$ in the diagram of $\tau-s$ is displayed in Fig. \ref{taus151}. $\tau_{max}$ is related with the annihilation point. As $\tau<\tau_{max}$, there are two intersection points for the AdS black hole under the $(5,1)$-model frame in the grand canonical ensemble. The red and blue lines are for the low-potential black hole (LPSB) and high-potential black hole (HPBH), respectively. They are just the stable and unstable black hole states. The corresponding winding numbers are 1, -1, so the topological number is $W=1-1=0$ as $\tau<\tau_{max}$. For $\tau=\tau_{max}$, the two intersection points for this type AdS black hole coincide. Furthermore, there is no black hole as $\tau>\tau_{max}$.

\subsection{Topology of RPST for $(6,2)$-model: BI gravity coupled to Maxwell's field in the canonical ensemble}
\label{section3.3}

When adopting $(d,k)=(6,2)$, the Wald entropy for the charged spherically symmetric AdS black hole solution was given in Ref. \cite{Kong2022} as
\begin{eqnarray}
S=\pi\sqrt{\frac{8Mr_h^3-2Q^2/\pi}{G_{(6,2)}l^4}}-\frac{r_h^4}{G_{(6,2)}l^2}
\end{eqnarray}
With the definitions $\bar S=L^2S/N+\pi l^2,~\bar Q=Q/N$, the corresponding radiation temperature and mass parameter are
\begin{eqnarray}
T&=&\frac{5\bar S^2-14\pi^{1/2}l\bar S^{3/2}+13\pi l^2\bar S-4\pi^{3/2}l^3\bar S^{1/2}-\pi^2L^2\bar Q^2}{8\pi^{5/4}l^{3/2}\bar S^{1/2}\left(\bar S^{1/2}-\pi^{1/2}l\right)^{5/2}},\label{T162}\\
M&=&N\frac{3\bar S^2-6\pi^{1/2}l\bar S^{3/2}+3\pi l^2\bar S+\pi^2L^2\bar Q^2}{6\pi^{5/4}l^{3/2}L^2\left(\bar S^{1/2}-\pi^{1/2}l\right)^{3/2}}.\label{M162}
\end{eqnarray}
From these thermodynamical parameters, the critical point reads
\begin{eqnarray}
\bar S_c&=&(0.9058+\pi)l^2,~\bar Q=0.0358l^2/L,~
T_c=0.1747/l,\\
\widetilde{\Phi}_c&=&0.2406/(lL),~
M_c=0.241lN/L^2,~F_c=0.0828lN/L^2.
\end{eqnarray}
\begin{figure}[htp]
\centering
\includegraphics[width=0.45\textwidth]{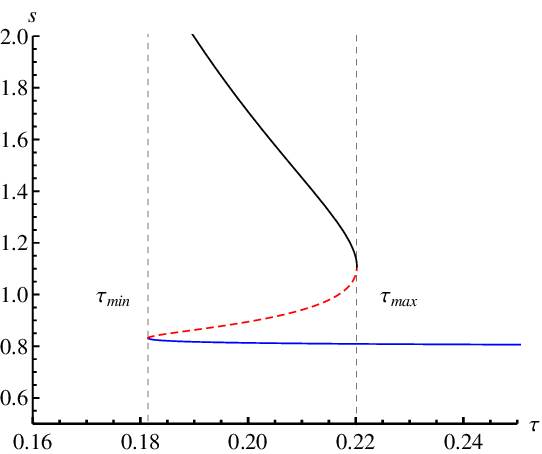}
\caption{Zero points of $\phi^s$ in the diagram of $\tau-s$ with $q=0.4$ for the higher dimensions and higher order curvature gravity model with $(d,k)=(6,2)$ in the canonical ensemble.}\label{taus62}
\end{figure}
By introducing the relative parameters $f=F/F_c=(M-TS)/F_{c},~1/\tau=t=T/T_{c},~s=\bar S/\bar S_{c},~q=\bar Q/\bar Q_c$, the component of the mapping $\phi^s$ reads
\begin{eqnarray}
\phi^s=\frac{\partial }{\partial s}\frac{3(0.9058+\pi)^2s^2
-6\sqrt{\pi}(0.9058+\pi)^{3/2}s^{3/2}
+3\pi(0.9058+\pi)s+(0.0358q\pi)^2}
{0.4968\pi^{5/4}\left[\sqrt{(0.9058+\pi)s}-\sqrt{\pi}\right]^{3/2}}
-\frac{1.91115}{\tau}.
\end{eqnarray}
The zero points of the mapping $\phi$ can be calculated by $\phi^s=0$. In the Fig. \ref{taus62}, we can see that there are three branches as $\tau_{min}<\tau<\tau_{max}$, the large black hole branch for $\tau<\tau_{min}$, and the small black hole branch for $\tau>\tau_{max}$. That means the existence of a phase transition in the region of $\tau_{min}<\tau<\tau_{max}$. The winding numbers of the large and small black hole branches, which are stable, both read $w=1$. And the corresponding topological number equals to one. While for the system in the region of $\tau_{min}<\tau<\tau_{max}$, there exist the stable large and small black hole, and the unstable intermediate black hole, thus the corresponding winding number are $w=1,1,-1$. The topological number is $W=1$. These results are consistent with that for the RN-AdS black hole and the AdS black hole under the $(5,1)$-model in the canonical ensemble.

\subsection{Topology of RPST for $(6,2)$-model: BI gravity coupled to Maxwell's field in the grand canonical ensemble}
\label{section3.4}

In this part we will consider the topology of the charged AdS black hole for the case of $(6,2)$-model in the grand canonical ensemble. From eq. (\ref{Sdk}), we have
\begin{eqnarray}
r_h^2=l^2\left(-1+\sqrt{1+\frac{SG_{(6,2)}}{\pi l^2}}\right).
\end{eqnarray}
With the above equation, eqs. (\ref{Nmu51}), (\ref{QPhi51}), and the definition $\widetilde{S}=\bar S/(\pi l^2)=\frac{L^2S}{\pi l^2N}+1,~\bar Q=Q/N$, the eqs. (\ref{T162}) and (\ref{M162}) can be rewritten as
\begin{eqnarray}
M&=&\frac{Nl\left[\widetilde{S}^2-2\widetilde{S}^{3/2}+\widetilde{S}
+3\widetilde{\Phi}^2N^2l^2L^2(\sqrt{\widetilde{S}}-1)^3\right]}
{2L^2(\sqrt{\widetilde{S}}-1)^{3/2}},\\
T&=&\frac{5\widetilde{S}^2-14\widetilde{S}^{3/2}+13\widetilde{S}
-4\widetilde{S}^{1/2}-9\widetilde{\Phi}^2N^2l^2L^2(\sqrt{\widetilde{S}}-1)^3}
{8\pi l\sqrt{\widetilde{S}}(\sqrt{\widetilde{S}}-1)^{5/2}}.
\end{eqnarray}
\begin{figure}[htp]
\centering
\includegraphics[width=0.45\textwidth]{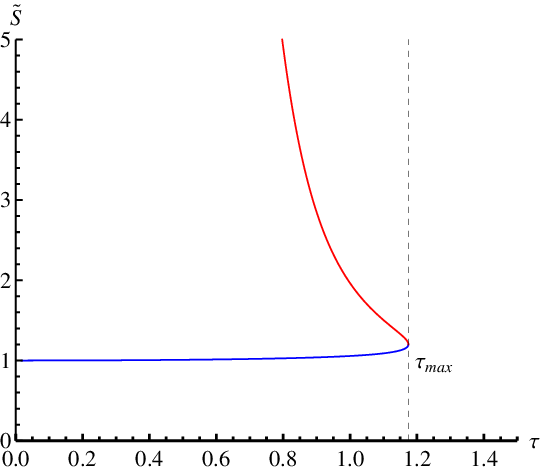}
\caption{Zero points of $\phi^{\widetilde{S}}$ in the diagram of $\tau-\widetilde{S}$ with $\widetilde{\Phi}^2N^2l^2L^2=0.8$ for the AdS black hole under the (6,2)-model frame in the grand canonical ensemble.}\label{taus162}
\end{figure}
Through the redefinitions $f(\widetilde{S})=F\L^2/(lN),~m(\widetilde{S})=M\L^2/(lN),
~1/\tau=t(\widetilde{S})=TlN$,
the components of the mapping $\phi$ read
\begin{eqnarray}
\phi^{\widetilde{S}}=\frac{\partial f(\widetilde{S})}{\partial\widetilde{S}}=\frac{\partial m(\widetilde{S})}{\partial\widetilde{S}}-\frac{1}{\tau},~~
\phi^\theta=-\cot\theta\csc\theta,
\end{eqnarray}
From the zero-point equation of the mapping $\phi^{\widetilde{S}}=0$, we can obtain
\begin{eqnarray}
\frac{1}{\tau}=\frac{5\widetilde{S}
+\sqrt{\widetilde{S}}(9\widetilde{\Phi}^2N^2l^2L^2-4)
-9\widetilde{\Phi}^2N^2l^2L^2}
{8\sqrt{\widetilde{S}}\sqrt{\sqrt{\widetilde{S}}-1}}.
\end{eqnarray}
The zero points of $\phi^{\widetilde{S}}$ in the diagram of $\tau-\widetilde{S}$ is displayed in Fig. \ref{taus162}. $\tau_{max}$ is related with the annihilation point. As $\tau<\tau_{max}$, there are two intersection points for the AdS black hole under the $(6,2)$-model frame in the grand canonical ensemble. The red and blue lines are for the low-potential black hole (LPSB) and high-potential black hole (HPBH), respectively. They are just the stable and unstable black hole states. The corresponding winding numbers are 1, -1, so the topological number is $W=1-1=0$ as $\tau<\tau_{max}$. For $\tau=\tau_{max}$, the two intersection points for this type AdS black hole coincide. Furthermore, there is no black hole as $\tau>\tau_{max}$.

\section{Discussions and conclusions}
\label{section4}

In this work, under the restricted phase space frame we investigated the thermodynamical topology of the RN-AdS black hole as well as two typical models of $(d,k)=(5,1),~(6,2)$ in the higher dimensions and higher order curvature gravity. The results showed that the topology numbers for these black holes in the canonical and grand canonical ensembles are different. For the charged AdS black holes (both the RN-AdS black hole and two others with $(d,k)=(5,1),~(6,2)$) in the canonical ensemble, the topological number is positive one, while it is zero for the charged AdS black holes in the grand canonical ensemble. This also proves that the EH and BI like models belong to the same gravitational classification from the thermodynamical topology perspective.

\section*{Acknowledgements}

We would like to thank Prof. Ren Zhao for his indispensable discussions and comments. This work was supported by the National Natural Science Foundation of China (Grant No. 12075143, Grant No. 12375050), and the Natural Science Foundation of Shanxi Province, China (Grant No. 202203021221209, Grant No. 202303021211180).


\begin{thebibliography}{99}
\bibitem{Bardeen1973}J. M. Bardeen, B. Carter, and S. W. Hawking, 
    Commun. Math. Phys. 31, 161 (1973).
\bibitem{Bekenstein1973}J. D. Bekenstein, 
    Phys. Rev. D 7, 949 (1973).
\bibitem{Hawking1983}S. W. Hawking and Don N. Page, 
    Commun. Math. Phys. 87, 577-588 (1983).
\bibitem{Witten1998}E. Witten, 
    Adv. Theor. Math. Phys. 2, 505 (1998), arXiv:hep-th/9803131.
\bibitem{Kastor2009}D. Kastor, S. Ray, and J. Traschen, 
    Class. Quant. Grav. 26 (2009) 195011, arXiv:0904.2765.
\bibitem{Hendi2017a}R. A. Hennigar, E. Tjoa, and R. B. Mann, 
    J. High Energ. Phys (2017) 70, arXiv:1612.06852.
\bibitem{Hennigar2017a}R. A. Hennigar and R. B. Mann, 
    Phys. Rev. Lett. 118, 021301 (2017), arXiv:1609.02564.
\bibitem{Frassin}A. M. Frassino, D. Kubiznak, R. B. Mann, et al., 
    J. High Energ. Phys, 09 (2014) 080, arXiv:1406.7015.
\bibitem{Kubiznak2012} D. Kubiznak and R. B. Mann, 
    J. High Energ. Phys 1207 (2012) 033, arXiv:1205.0559.
\bibitem{Cai2013}R.-G Cai, L.-M Cao, L. Li, et al., 
    J. High Energ. Phys, 2013 (9), arXiv:1306.6233.
\bibitem{Ma2017}M.-S Ma, R. Zhao, and Y.-S Liu, 
    Class. Quant. Grav. 34 (2017) 165009, arXiv:1604.06998.
\bibitem{Banerjee2017}R. Banerjee, B. R. Majhi, and S. Samanta, 
    Phys. Lett. B 767 (2017) 25-28, arXiv:1611.06701.
\bibitem{Mann1207}D. Kubiznak and R. B. Mann, 
    J. High Energ. Phys 1207 (2012) 033, arXiv:1205.0559.
\bibitem{Wei2015}S.-W Wei and Y.-X Liu, 
    Phys. Rev. Lett 115 (2015) 111302.
\bibitem{Bhattacharya2017}K. Bhattacharya, B. R. Majhi, and S. Samanta, 
    Phys. Rev. D 96, 084037 (2017).
\bibitem{Zeng2017}X.-X Zeng and L.-F Li, 
    Phys. Lett. B 764 (2017) 100.
\bibitem{Zhang1502}J.-L Zhang, R.-G Cai, and H.-W Yu, 
\bibitem{Du2021}Y.-Z Du, H.-F Li, F. Liu, et al.,  
    Chin. Phys. C 45 (2021) 11, arXiv:2112.10403.
\bibitem{Zhang2020}Y. Zhang, W.-Q Wang, Y.-B Ma, et al., 
    Adv. H. E. Phys. 2020 (2020) 7263059, arXiv:2004.06796.
\bibitem{Visser2022}M. R. Visser, 
    Phys. Rev. D 105, 106014 (2022).
\bibitem{Ahmed2023}M. B. Ahmed, W. Cong, D. Kubiznak, et al., 
    Phys. Rev. Lett. 130, 181401 (2023).
\bibitem{Zhao2022}L. Zhao, 
    Chin. Phys. C 46, 055105 (2022).
\bibitem{Gao2022}Z.-Y Gao, X. Kong, and L. Zhao, 
    Eur. Phys. J. C 82, 112 (2022).
\bibitem{Gao2022a}Z. Y. Gao and L. Zhao, 
    Class. Quant. Grav. 39, 075019 (2022).
\bibitem{Du2023}Y.-Z Du, H.-F Li, Y. Zhang, et al., 
    Entropy 25 (2023) 4, arXiv:2210.02006.
\bibitem{Duan1979}Y.-S Duan and M. L. Ge, 
    Sci. Sin. 9, 1072 (1979).
\bibitem{Wei2022}S.-W Wei, Y.-X Liu, and R. B. Mann, 
    Phys. Rev. L 129, 191101 (2022).
\bibitem{Kong2022}X.-Q Kong, T Wang, Z.-Y Gao, et al., 
    Entropy 24 (2022) 8, arXiv:2208.07748.
\bibitem{Cai2002}R.-G Cai, 
    Phys. Rev. D, 2002, 65 (8):084014, arXiv:hepth/0109133.
\bibitem{Lanczos1932}C. Lanczos,
    Z. Phys., 1932, 73(3):147-168.
\bibitem{Lovelock1971}D. Lovelock, 
    J. Math. Phys., 1971, 12(3):498-501.

\bibitem{Wei2020}S.-W. Wei, 
    Phys. Rev. D 102, 064039 (2020).

\bibitem{Wei2022a}S.-W. Wei and Y.-X. Liu, 
    Phys. Rev. D 105, 104003 (2022).
\bibitem{Wu2023}D Wu, 
    Eur. Phys. J. C 83 (2023) 589,arXiv:2306.02324[gr-qc]
\bibitem{Wu2023a}D Wu, 
    Eur. Phys. J. C 83 (2023) 365, arXiv:2302.01100.

\bibitem{Duan1984} Y.-S. Duan, {\it The structure of the topological current,}  Report No. SLAC-PUB-3301 (1984).
\bibitem{Duan1998}Y.-S. Duan, S. Li, and G.-H. Yang, 
    Nucl. Phys. B514, 705 (1998).
\bibitem{Fu2000}L.-B. Fu, Y.-S. Duan, and H. Zhang, 
    Phys. Rev. D 61, 045004 (2000).
\bibitem{Fan2023}Z.-Y. Fan, 
    Phys. Rev. D 107, 044026 (2023).


\bibitem{Gogoi2023}N. J. Gogoi and P. Phukon, 
    Phys. Rev. D 108 (2023) 6, arXiv:2304.05695.
\bibitem{Crisostomo2000}J. Crisostomo, R. Troncoso, and J. Zanelli, 
    Phys. Rev. D, 62 (2000) 8, arXiv:hep-th/0003271.
\bibitem{Banados1992}M. Banados, C. Teitelboim, and J. Zanelli, 
    Phys. Rev. Lett. 69 (1992) 1849, arXiv:hep-th/9204099.
\bibitem{Banados1993}M. Banados, M. Henneaux, C. Teitelboim, et al., 
    Phys. Rev. D 48 (1993) 1506, arXiv:gr-qc/9302012.
\bibitem{Banados1994}M. Banados, C. Teitelboim, and J. Zanelli, 
    Phys. Rev. D 49 (1994) 975.
\end{thebibliography}
\end{document}